\documentclass[conference,10pt]{IEEEtran}
\usepackage{epsfig,rotating,setspace,latexsym,amsmath,epsf,amssymb,amsfonts,bm,theorem,subfigure,epstopdf}
\usepackage{cite,authblk}
\usepackage{bbm}
\usepackage{algorithmic,algorithm}

\IEEEoverridecommandlockouts

\allowdisplaybreaks
\usepackage{color}

\begin{document}

\title{Coded Status Updates in an Energy Harvesting Erasure Channel
	\thanks{This work was supported by NSF Grants CNS 13-14733, CCF 14-22111, CNS 15-26608, and CCF 17-13977.}}

\author{Abdulrahman Baknina \qquad Sennur Ulukus\\
	\normalsize Department of Electrical and Computer Engineering\\
	\normalsize University of Maryland, College Park, MD 20742\\
	\normalsize  \emph{abaknina@umd.edu} \qquad \emph{ulukus@umd.edu}}

\maketitle

\begin{abstract}
	We consider an energy harvesting transmitter sending status updates to a receiver over an erasure channel, where each status update is of length $k$ symbols. The energy arrivals and the channel erasures are independent and identically distributed (i.i.d.) and Bernoulli distributed in each slot. 
	In order to combat the effects of the erasures in the channel and the uncertainty in the energy arrivals, we use channel coding to encode the status update symbols.
	 We consider two types of channel coding: maximum distance separable (MDS) codes and rateless erasure codes.
	For each of these models, we study two achievable schemes: best-effort and save-and-transmit. 
	In the best-effort scheme, the transmitter starts transmission right away, and sends a symbol if it has energy. 
	In the save-and-transmit scheme, the transmitter remains silent in the beginning in order to save some energy to minimize energy outages in future slots. We analyze the average age of information (AoI) under each of these policies. We show through numerical results that as the average recharge rate decreases, MDS coding with save-and-transmit outperforms all best-effort schemes. We show that rateless coding with save-and-transmit outperforms all the other schemes. 
\end{abstract}

\section{Introduction}
We consider an energy harvesting single-user system, where the communication channel between the transmitter and the receiver is an erasure channel. The transmitter collects measurements of a certain phenomenon and sends updates on this phenomenon to the receiver; these updates are referred to as \emph{status updates}. The purpose of sending status updates is to minimize the age of information (AoI) at the receiver.

Energy harvesting communications with the objective of maximizing the throughput has been extensively studied, for example, see \cite{jingP2P, kayaEmax, omurFade, ruiZhangEH,jingBC,jingMAC, aggarwalPmax,orhan2015energy,gurakan2016cooperative,gunduz2hop,varan_twc_jour, arafa2017energy,gunduzLoss, orhan-broadband, ruiZhangNonIdeal, omurHybrid, kayaLoss,yatesRxEH1, yatesRxEH2, yatesRxEH3, payaroRxEH, arafaJSACdec,omur_temp_journ, bakninaenergy,BOU-gc17}. 
The single-user channel is studied in \cite{jingP2P, kayaEmax, omurFade, ruiZhangEH}, extended to multi-user settings in \cite{jingBC,jingMAC, aggarwalPmax}, multi-hop channels in \cite{orhan2015energy,gurakan2016cooperative,gunduz2hop}, and two-way channels in \cite{varan_twc_jour,arafa2017energy}. 
Effects of imperfect circuitry, receiver side processing, and temperature increases are considered in \cite{gunduzLoss, orhan-broadband, ruiZhangNonIdeal, omurHybrid, kayaLoss,yatesRxEH1, yatesRxEH2, yatesRxEH3, payaroRxEH, arafaJSACdec,omur_temp_journ, bakninaenergy,BOU-gc17}.

 \begin{figure}[t]
	\centerline{\includegraphics[width=0.9\columnwidth]{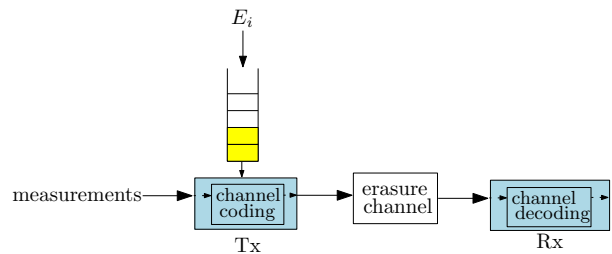}}
	\vspace{-0.2cm}
	\caption{An energy harvesting transmitter with an infinite battery. The transmitter collects measurements and sends updates to the receiver over an erasure channel.}
	\label{sys_model}
	\vspace{-0.5cm}
\end{figure}

In this paper, we consider an energy harvesting communication system with the objective of minimizing the average AoI at the receiver. Status updates and AoI metric is studied in many different settings, for example, see \cite{kaul2012real,kaul2012status,yates2012real,kam2013age,costa2014age,sun2016update,kosta2017age,bedewy2017age,parag2017real,yates2017timely,yates2015lazy,bacinoglu2015age,wu2017optimal,arafa2017age2,arafa2017age}. 
References \cite{kaul2012real,kaul2012status,yates2012real,kam2013age,costa2014age} study minimizing the AoI with a queuing theoretic approach; penalty functions and non-linear costs are studied in \cite{sun2016update,kosta2017age}; the optimality of last-come-first-serve for multi-hop settings is shown in \cite{bedewy2017age}; and erasure channels are considered in \cite{parag2017real,yates2017timely}. The energy harvesting case and when the energy arrivals are known only causally is studied in \cite{yates2015lazy,bacinoglu2015age,wu2017optimal}. The optimality of threshold policies for the case of unit batteries is shown in \cite{wu2017optimal}. Energy harvesting single-user and multi-hop settings with non-causal energy arrival knowledge are studied in \cite{arafa2017age2,arafa2017age}. 

This paper is closely related to \cite{yates2017timely}, in which coded status updates are proposed in order to overcome channel errors. We consider a single-user channel shown in Fig. \ref{sys_model}, where the transmitter is energy harvesting and further transmission errors may occur due to energy outages. We consider two different types of channel codes to encode the status updates. First, we consider maximum distance separable (MDS) codes. With MDS coding, the transmitter encodes the $k$ status update symbols into $n$ symbols. The receiver receives the update successfully if it receives any $k$ of these $n$ encoded symbols. 
Next, we consider rateless codes, for example, fountain codes. In this case, the transmitter encodes the $k$ update symbols into as many symbols as needed until $k$ of these symbols are received successfully.
For each of these models, we consider two different policies: best-effort and save-and-transmit. Best-effort and save-and-transmit schemes were originally considered in \cite{ozel2012achieving}, in the context of achieving the capacity of the energy harvesting AWGN channel. In the best-effort scheme, in each slot, the transmitted symbol may suffer from two errors: channel erasure and energy outage. In the save-and-transmit scheme, the transmitter remains silent at the beginning to save energy and to reduce the errors due to energy outage. 

For all these cases, we derive the average AoI. Through numerical results, we show that as the average recharge rate decreases, MDS codes with save-and-transmit outperforms all the best-effort schemes. The gain becomes significant for low values of average energy arrivals. We observe that rateless coding with save-and-transmit outperforms all other policies.

\section{System Model}\label{sec_sys_model} 
We consider a single-user channel with a transmitter which has an infinite-sized battery, see Fig. \ref{sys_model}. The energy arrivals are Bernoulli and i.i.d.: in slot $i$, a unit energy arrives with probability $p$ or no energy arrives with probability $1-p$, i.e., $\mathbb{P}[E_i=1]=1-\mathbb{P}[E_i=0]=p$. The transmitter obtains the measurements (status updates), which are packets of length $k$, which should be sent to the receiver in a way to minimize the average AoI at the receiver. 

The total AoI up to time $T$ is,
\begin{align}
\Delta_T= \int_{0}^{T} \left(t-u(t)\right) dt
\end{align}
where $u(t)$ is the time stamp of the latest received status update packet and $\Delta(t)=t-u(t)$ is the instantaneous AoI.

\begin{figure}[t]
	\centerline{\includegraphics[width=0.9\columnwidth]{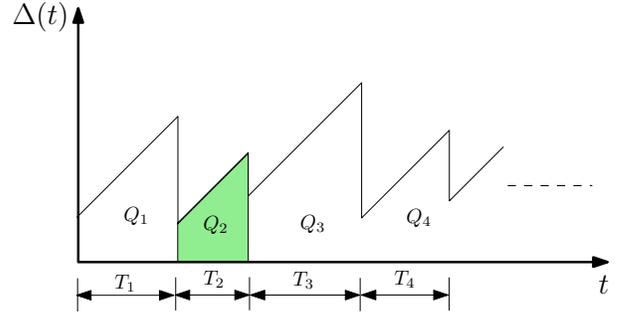}}
	\vspace{-0.2cm}
	\caption{An example for the evolution of the age of information.}
	\label{AoI_general}
	\vspace{-0.5cm}
\end{figure}

An example evolution of the AoI is shown in Fig. \ref{AoI_general}. The average long-term AoI in this case is calculated as,
\begin{align}
\Delta = \lim_{T\rightarrow \infty} \frac{\Delta_T}{T} = \lim_{i\rightarrow \infty} \frac{\sum_{j=1}^{i}Q_j}{\sum_{j=1}^{i}T_j}
\end{align}
In all the subsequent analysis we will assume renewal policies, i.e., where $Q_j$ and $T_j$ are i.i.d. The AoI then reduces to,
\begin{align}
\Delta = \lim_{i\rightarrow \infty} \frac{\frac{1}{i}\sum_{j=1}^{i}Q_j}{\frac{1}{i}\sum_{j=1}^{i}T_j} =\frac{\mathbb{E}[Q]}{ \mathbb{E}[T]}
\end{align}
where we dropped the subscript $j$ as $Q_j$ and $T_j$ are i.i.d.

The channel between the transmitter and the receiver is an i.i.d. erasure channel. The probability of symbol erasure (loss) in each slot is $\delta$. In order to combat the channel erasures and the energy outages, the transmitter encodes the status updates before sending them through the channel.

We consider two types of channel codes: MDS and rateless codes. We first consider MDS channel codes. For this case we have an $(n,k)$ channel coding scheme, where $k$ is the length of an uncoded status update and $n$ is the length of an encoded codeword which is sent through the channel with $n \geq k$. When the transmitter is done with sending the $n$ symbols, it generates a new update and begins sending it. This is irrespective of the success of the transmission of these $n$ symbols. The optimal value of $n$ depends on $k$, $\delta$, and $p$. For MDS channel coding, we study two achievable schemes. We first study a save-and-transmit scheme in which the transmitter saves energy from the incoming energy arrivals until it has at least $n$ units of energy in its battery. This in effect makes sure that errors which can occur during the codeword transmission are only due to the erasures in the channel. To ensure that the synchronization is maintained between the transmitter and the receiver, the transmitter remains in the saving phase for a number of slots which is multiple of $n$. We then study a best-effort scheme, in which the transmitter attempts transmission in each slot. In this case, the error in each symbol can be either due to an energy outage or a channel erasure or both. 

We next study the case of rateless coding in which the transmitter keeps sending the update until $k$ symbols are successfully received. For this case, we also study two schemes: best-effort and save-and-transmit. In the best-effort scheme, once the update is successfully received, the transmitter generates a new update and begins transmitting it immediately. In the save-and-transmit scheme, once the update is successfully received, the transmitter waits some time in order to save some energy in the battery to prevent future energy outages. The transmitter saves for $m$ slots, where the optimal $m$ should be obtained as a function of the system parameters $\delta$, $k$, and $p$.

%

\section{AoI Under MDS Channel Coding}

\subsection{Save-and-Transmit Policy}
In the save-and-transmit policy, before the transmitter attempts to transmit the coded update, the transmitter remains silent for an integer multiple of $n$ slots until the battery has energy at least equal to $n$. The duration the transmitter remains silent for the $j$th time while transmitting the $i$th update is a random variable denoted by $Z_{ij} \in \{n,2n,3n,\ldots \}$ which depends on the energy arrival distribution. 
The random variable $Z_{ij}$ can be expressed as:
\begin{align}
Z_{ij}=\left\lceil \frac{W_i}{n} \right\rceil n
\end{align}
where $W_i$ is the random variable which denotes the number of slots needed to save $n$ units of energy and $\lceil x \rceil$ denotes the smallest integer greater than or equal to $x$. 
Since the energy arrivals follow an i.i.d. Bernoulli distribution, $W_i$ will follow a negative binomial distribution as follows:
\begin{align}
P_{W_i}(w)=   {{w-1}\choose{n-1}} p^n (1-p)^{w-n}, \ \ w=n,n+1,\ldots \label{eq_neg_binomial}
\end{align}
The distribution of $Z_{ij}$ can be obtained using (\ref{eq_neg_binomial}) as follows:
\begin{align}
P_{Z_{ij}}(z) &= \sum_{w=z-n+1}^{z} P_{W_i}(w) , \ \ z=n,2n,\ldots \label{pmf_Z_12}
\end{align}

After the saving phase, the transmission resumes for $n$ slots. After the transmitter is done transmitting the $n$ coded symbols, the transmitter again goes to the saving phase until it recharges its battery to at least $n$. The transmitter alternates between saving and transmission phases. 

The update is successful if at least $k$ symbols are received without being erased; there will be no energy outage due to the saving phase. Hence, the probability of having a success in a $n$ slot of duration is,
\begin{align}
\epsilon_{k,n}(\delta) = \sum_{x=k}^{n} {{x-1}\choose{k-1}} (1-\delta)^k \delta^{x-k}
\end{align}
Thus, in the $n$ consecutive slots the transmission is successful with probability $\epsilon_{k,n}(\delta)$. Now, the update will be successful in the $V$th transmission, where $V$ is a geometrically distributed random variable with a the following pmf,
\begin{align}
P_{V^{(n)}}(v)=\epsilon_{k,n}(\delta) (1-\epsilon_{k,n}(\delta))^{v-1}, \ \ v=1,2,\ldots
\end{align}
Hence, we may need to repeat the save-and-transmit phases for $V$ times before we have a successful status update.

We now characterize the random variable which identifies the instant at which the update will be successful within the $n$ consecutive slots. We denote this random variable by $\tilde{X}_i$ which has a conditional pmf $P_{X_i|X_i \leq n}(x)$ where
\begin{align}
P_{X_i}(x)=   {{x-1}\choose{k-1}} (1-\delta)^k \delta^{x-k}, \ \ x=k,k+1,\ldots
\end{align}
Hence, $\tilde{X}_i$ is distributed as:
\begin{align}
P_{\tilde{X}_i}(x)=\frac{{{x-1}\choose{k-1}} (1-\delta)^k \delta^{x-k}}{\epsilon_{k,n}(\delta)}, \ \ x=k,k+1, \ldots,n \label{pmf_x_tilde}
\end{align}

\begin{figure}[t]
	\centerline{\includegraphics[width=0.9\columnwidth]{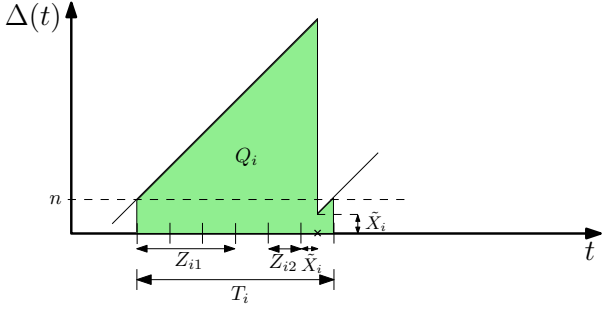}}
	\vspace{-0.2cm}
	\caption{An example for the evolution of the age of information under the save-and-transmit scheme for the MDS channel coding case.}
	\label{AoI_FR_ST}
	\vspace{-0.5cm}
\end{figure}

An example which illustrates the AoI evolution is shown in Fig. \ref{AoI_FR_ST}. In this figure, the transmitter at first waits $3n$ slots in order to recharge the battery to at least the level $n$. It then attempts to transmit. The transmission in this case is not successful due to the channel erasures so the transmitter again waits for $n$ slots in order to charge the battery. The transmission then proceeds again in the next slot. The transmission is then successful and the receiver received the update after $\tilde{X}_i$ transmissions, where $k \leq \tilde{X}_i \leq n$.

We now consider a renewal policy which serves as an upper bound for the save-and-transmit policy described above. We assume that at the end of the update period, the transmitter depletes all its battery. Thus, the transmitter renews its state at the end of each successful update and always begins with a depleted battery. In this case, the AoI can be written as:
\begin{align}
\Delta_{MDS-ST}= \frac{\mathbb{E}[Q_i]}{\mathbb{E}[T_i]}\label{AoI_BC-ST}
\end{align}
Next, we evaluate $\mathbb{E}[Q_i]$ and $\mathbb{E}[T_i]$. We first obtain $Q_i$ as,
\begin{align}
Q_i=& n\left[n \left(V_i-1 \right) + \tilde{X}_i + \sum_{j=1}^{V_i} Z_{ij}\right] \nonumber \\
& +\frac{1}{2}\left[n \left(V_i-1 \right) + \tilde{X}_i + \sum_{j=1}^{V_i} Z_{ij}\right]^2 + \frac{n^2}{2} - \frac{\tilde{X}_i^2}{2} \\
=& n^2 \frac{{V_i}^2}{2}+ n V_i \tilde{X}_i + n \sum_{j=1}^{V_i} Z_{ij} \nonumber \\
&+\left[n \left( V_i-1\right) + \tilde{X}_i\right] \sum_{j=1}^{V_i} Z_{ij} + \frac{1}{2}\left(\sum_{j=1}^{V_i} Z_{ij} \right)^2
\end{align}
We then obtain $T_i$ as,
\begin{align}
T_i=n V_i+\sum_{j=1}^{V_i} Z_{ij}
\end{align}

Now, it remains to calculate the expectation of $Q_i$ and $T_i$. We first calculate the first and second moments of $\sum_{j=1}^{V_i} Z_{ij}$, using \cite[Theorem 6.13]{yates1999probability}, as follows:
\begin{align}
\mathbb{E} \left[\sum_{j=1}^{V_i} Z_{ij}\right] 
=& \frac{\mathbb{E} \left[  Z \right]}{\epsilon_{k,n}(\delta)} 
\end{align}
Similarly, we have:
\begin{align}
\mathbb{E}\left[\left(\sum_{j=1}^{V_i} Z_{ij} \right)^2\right] 
=& \frac{\mathbb{E} \left[  Z^2   \right]}{\epsilon_{k,n}(\delta)} +
\frac{2-2\epsilon_{k,n}(\delta)}{\epsilon_{k,n}^2(\delta)}  \mathbb{E} \left[  Z \right]^2
\end{align}
We then combine all these to obtain:
\begin{align}
\mathbb{E}\left[T_i\right]=\frac{n}{\epsilon_{k,n}(\delta)} +  \frac{\mathbb{E} \left[  Z \right]}{\epsilon_{k,n}(\delta)}  \label{T_i}
\end{align}
and 
\begin{align}
\mathbb{E}\left[Q_i\right]=& \frac{n^2(2-\epsilon_{k,n}(\delta))}{2 \epsilon_{k,n}^2(\delta)} + \frac{n\mu_{\tilde{X}}}{\epsilon_{k,n}(\delta)} +  \frac{n (2-\epsilon_{k,n}(\delta)) \mathbb{E} \left[  Z \right] }{ \epsilon_{k,n}^2(\delta)} \nonumber \\
&+\! \frac{ \mu_{\tilde{X}}\mathbb{E} \left[  Z \right] }{\epsilon_{k,n}(\delta)} \!+\! \frac{1}{2} \frac{ \mathbb{E} \left[  Z^2 \right] }{\epsilon_{k,n}(\delta)} \!+\! \frac{ (1\!-\!\epsilon_{k,n}(\delta)) \mathbb{E} \left[  Z \right]^2 }{\epsilon_{k,n}^2(\delta)} \label{Qi}
\end{align} 
where $\mathbb{E} \left[  Z \right]$ and $\mathbb{E} \left[  Z^2 \right]$ can be calculated using (\ref{pmf_Z_12}) and $\mu_{\tilde{X}}$ can be calculated using (\ref{pmf_x_tilde}).
Hence, the average AoI $\Delta_{MDS-ST}$ in (\ref{AoI_BC-ST}) can be found by substituting with the expressions in (\ref{T_i}) and (\ref{Qi}).

\subsection{Best-Effort Policy}
We now consider the case when the transmitter does not wait at the beginning in order to save energy, instead it begins transmission immediately. The error events in this case can be either an erasure in the communication channel or an energy outage at the transmitter. These two events may occur for each transmitted symbol. Hence, for the symbol to be received without an error, there should be no energy outage and no channel erasure; this forms a Bernoulli random variable with probability of success equal to $q\triangleq p(1-\delta)$.
The evolution of AoI is similar to Fig. \ref{AoI_FR_ST} but in this case, $Z_{ij}$ is equal to zero as the transmitter does not wait to save energy.

Using analysis similar to the previous scheme, but with having the probability of success equal to $q$, the average AoI in this case can be written as:
\begin{align}
\Delta_{MDS-BE} = \frac{n}{\epsilon_n} - \frac{n}{2} +\frac{k \epsilon_{k+1,n+1}(q)}{q \epsilon_{k,n}(q)}
\end{align}
This can also be obtained using the same analysis as in \cite{yates2017timely}, but with probability of success equal to $q$,

\section{AoI Under Rateless Channel Coding} 

\subsection{Best-Effort Policy}\label{sub_sec_BE_RL}
We consider here the case when the transmitter begins to transmit immediately. In each slot, the transmitter suffers two possible error events. The first is channel erasure and the second is energy outage. Hence, a symbol will be received successfully if neither error occurs, which happens with probability equal to $q$. The channel is now equivalent to an erasure channel, similar to the one considered in \cite{yates2017timely}, but with probability of success equal to $q$. Following analysis similar to the one in \cite{yates2017timely}, but with probability of success equal to $q$, the average AoI in this case is equal to:
\begin{align}
\Delta_{RC-BE} =\frac{k}{q} \left(\frac{3}{2}+\frac{1-q}{k}\right)
\end{align}

%
%
%
%
%
%

\begin{figure}[t]
	\centerline{\includegraphics[width=1.0\columnwidth]{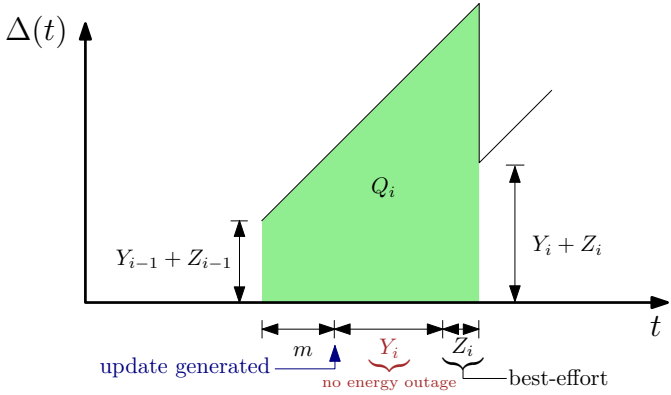}}
	\vspace{-0.2cm}
	\caption{An example for the evolution of the age of information under the save-and-transmit scheme for the rateless channel coding case.}
	\label{AoI_IIR_ST}
	\vspace{-0.4cm}
\end{figure}

\subsection{Save-and-Transmit Policy}
In this policy, we consider the case when the transmitter does not generate a new update immediately once the transmission of the previous update is successful, but it waits for a deterministic time of $m$ slots. Here, $m$ is a deterministic number which both the transmitter and the receiver know in advance; this $m$ should then be optimized to minimize the average AoI and will be a function of $\delta$, $p$ and $k$.

The transmission in this policy proceeds as follows: once the previous update is successful, the transmitter begins a saving phase of duration $m$ slots. Then, the transmitter generates a new update and begins transmitting it to the receiver. While transmitting the update, the transmitter may receive more energy arrivals; however, the amount of energy in the battery will always be non-increasing as the transmitter transmits a symbol in each slot while the energy may not arrive at every slot. The transmitter keeps transmitting the update until its battery state hits zero; this declares the end of the \emph{no-outage} phase. We denote the number of symbols sent successfully in this phase by $k_i$. If $k_i \geq k$, then no more transmission is required and the update is successful. Otherwise, the transmitter transmits the remaining $k-k_i$ using the best-effort scheme described in Subsection \ref{sub_sec_BE_RL}.

We denote the duration the transmitter transmits with no outage by $Y_i$ and we denote the duration we transmit using the best-effort scheme by $Z_i$. An example for the evolution of the AoI in this case is shown in Fig.~\ref{AoI_IIR_ST}. 
The average AoI can be calculate as follows,
\begin{align}
\Delta&_{RC-ST}= \frac{\mathbb{E}[Q_i]}{m+\mathbb{E}[Y_i+Z_i]} \\
=& \frac{\mathbb{E} \left[ \left(m+Y_i \!+\!Z_i\right)^2 \!+\! 2 \left(m\!+\!Y_i\!+\!Z_i\right) \left(Y_{i-1}\!+\!Z_{i-1}\right) \right]}{2m+2\mathbb{E}[Y_i+Z_i]}
\end{align}
This AoI can be calculated explicitly once $\mathbb{E}[Y_i]$, $\mathbb{E}[Y^2_i]$, $\mathbb{E}[Z_i]$, $\mathbb{E}[Z_i^2]$ and $\mathbb{E}[Y_i Z_i]$ are calculated. 
We note that $Y_i$ and $Z_i$ are dependent on each other while $Y_i$ and $Y_{i-1}$ are independent due to using a renewal policy. 

We now define the random variables $\{E_i\}_{i=1}^\infty$; the random variable $E_1$ represents the amount of energy harvested in the first $m$ slots. For $i\geq 2$, the random variable $E_i$ represents the amount of energy harvested during the previous $E_{i-1}$ slots. Hence, we have $E_i \leq E_{i-1}$.

\begin{figure}[t]
	\centerline{\includegraphics[width=0.9\columnwidth]{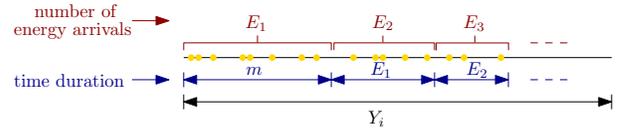}}
	\vspace{-0.2cm}
	\caption{An example to illustrate the random variable $Y_i$.}
	\label{AoI_IIR_ST_Y}
	\vspace{-0.5cm}
\end{figure}

We now characterize the random variable $Y_i$, 
\begin{align}
Y_i = \sum_{i=1}^{\infty} E_i
\end{align}
where $E_1$ is Bin$(m,p)$, and for $i\geq 2$, $E_i$ given $E_{i-1}=e_{i-1}$ is Bin$(e_{i-1},p)$; Bin$(.)$ denotes binomial distribution.
An example for the evolution of $Y_i$ is shown in Fig. \ref{AoI_IIR_ST_Y}.

We can obtain the marginal pmf for the random variables $E_i$, $i\geq2$, by applying \cite[Theorem 6.12]{yates1999probability} and using \cite[Table 6.1]{yates1999probability}. Each $E_i$ consists of a sum of i.i.d. Bernoulli random variables and the number of these random variables is distributed according to a binomial distribution of $E_{i-1}$ which is independent of the Bernoulli random variables. Hence, the marginal pmf of the random variable $E_i$ is Bin($m$,$p^i$).

We can now calculate $\mathbb{E}[Y_i]$ as,
\begin{align}
\mathbb{E}[Y_i]=&  \sum_{i=1}^{\infty} \mathbb{E}\left[  E_i \right] = \frac{mp}{1-p }
\end{align}
Next, we want to calculate $\mathbb{E}[Y_i^2]$ which we calculate as $\mathbb{E}[Y_i^2]= \mbox{var}(Y_i)+ \mathbb{E}[Y_i]^2 $. The term $\mbox{var}(Y_i)$ can be calculated as follows
\begin{align}
\mbox{var}(Y_i) =& \sum_{i=1}^{\infty} \mbox{var}(E_i) +2 \sum_{i<j}^{\infty} \mbox{cov}(E_i,E_j) \\
=& \frac{mp}{1-p^2} +2 \sum_{i<j} \mbox{cov}(E_i,E_j)
\end{align}
This requires the calculation of $\mbox{cov}(E_i,E_j)$, $\forall i>j$. To calculate the covariance, we first calculate the conditional probability $\mathbb{P}(E_{j+1}| E_i)$. 
For $j>i$, we have that $\mathbb{P}(E_{j}| E_i)$ is distributed as Bin($E_i$,$p^{j-i}$).
This again follows by applying \cite[Theorem 6.12]{yates1999probability} and using \cite[Table 6.1]{yates1999probability}.

We now calculate for $j>i$ $\mbox{cov}(E_j,E_i)$ as follows:
\begin{align}
\mbox{cov}(E_j,E_i)= &\mathbb{E}[E_j E_i] -\mathbb{E}[E_j]\mathbb{E}[E_i]
= m p^j(1-p^i)
\end{align}
Next, we calculate $\sum_{i<j} \mbox{cov}(E_i,E_j)$ as follows:
\begin{align}
\sum_{i<j} \mbox{cov}(E_i,E_j)=& \sum_{i=1}^{\infty} \sum_{j=i+1}^{\infty} m p^j(1-p^i) \\
=& \frac{mp^2}{(1-p)(1-p^2)} 
\end{align}
Therefore, $\mbox{var}(Y_i)$ is equal to
\begin{align}
\mbox{var}(Y_i)=& \frac{mp}{1\!-\!p^2} +2 \frac{mp^2}{(1\!-\!p)(1\!-\!p^2)} 
= \frac{mp(1\!+\!p)}{(1\!-\!p)(1\!-\!p^2)}
\end{align}
Hence, $\mathbb{E}[Y_i^2]$ can be calculated as follows:
\begin{align}
\mathbb{E}[Y_i^2]= \frac{mp(1+p)}{(1-p)(1-p^2)}+ \frac{m^2p^2}{(1-p)^2 }
\end{align}

Next, we calculate $\mathbb{E}[Z_i]$, $\mathbb{E}[Z_i^2]$ and $\mathbb{E}[Y_i Z_i]$. The pmf of $Z_i|Y_i=k_1$ is negative binomial distribution as in (\ref{eq_neg_binomial}) but with number of successes equal to $\max(k-k_1,0)$ and with success probability equal to $q$. The value of $\mathbb{E}[Z_i|Y_i=y_i]$ can then be calculated using conditional expectation as follows:
\begin{align}
\mathbb{E}[Z_i|Y_i=y_i]= \sum_{w=0}^{y_i} {{y_i}\choose{w}} \delta^{y_i-w} (1-\delta)^{w} \frac{g(w)}{q}
\end{align}
and the value of $\mathbb{E}[Z_i^2|Y_i=y_i]$ can be calculated as follows
\begin{align}
\!\!\! \! \mathbb{E}[Z_i^2|Y_i\!=\!y_i]\!\!=\!\!\! \sum_{w=0}^{y_i} \!\! {{y_i}\choose{w}}\delta^{y_i\!-\!w} (1\!\!-\!\!\delta)^{\!w}  \frac{g(w)(g(w)\!+\!(1\!\!-\!q))}{q^2} 
\end{align}
where $g(w)\triangleq\max(k-w,0)$. Similarly, we can obtain $\mathbb{E}[Y_i Z_i|Y_i=y_i]$. 
Now, it remains to calculate the expectation over the pmf of $Y_i$. Due to the dependency between the terms $E_i$ and their infinite sum, there is no closed form for the pmf of $Y_i$ and it can be found numerically.

\begin{figure}[t]
	\centerline{\includegraphics[width=0.9\columnwidth]{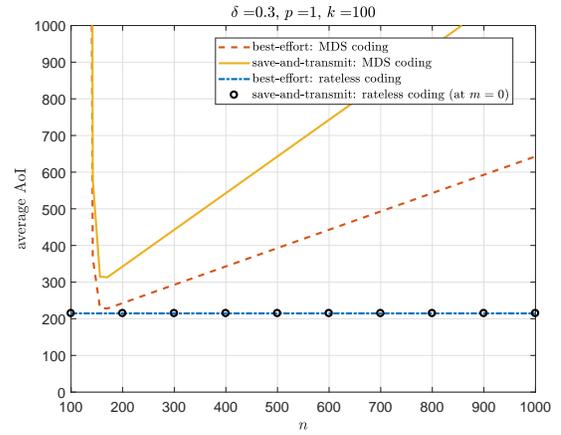}}
	\vspace{-0.5cm}
	\caption{Comparison of average AoI, $p=1$.}
	\label{res_4}
	\vspace{-0.5cm}
\end{figure}

\begin{figure}[t]
	\centerline{\includegraphics[width=0.9\columnwidth]{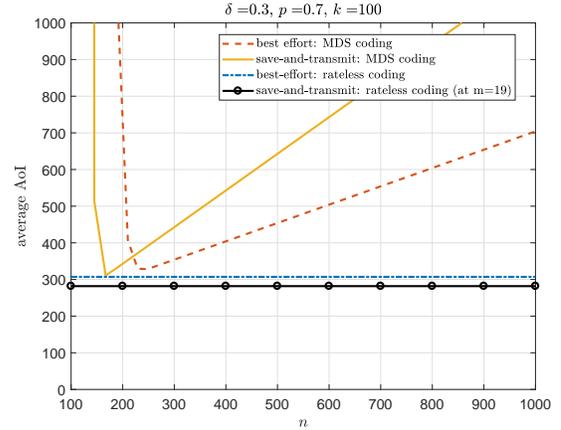}}
	\vspace{-0.5cm}
	\caption{Comparison of average AoI, $p=0.7$.}
	\label{res_1}
	\vspace{-0.6cm}
\end{figure}

\section{Numerical results}
\begin{figure}[t]
	\centerline{\includegraphics[width=0.9\columnwidth]{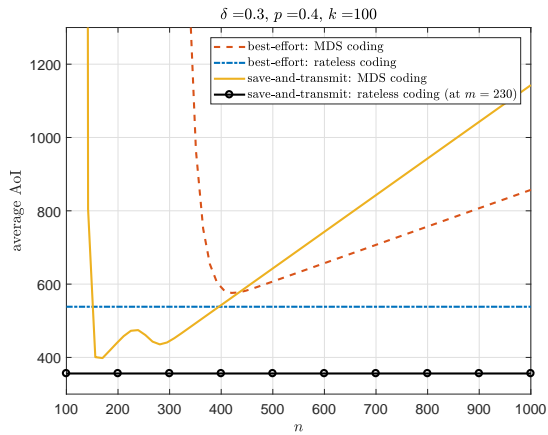}}
	\vspace{-0.5cm}
	\caption{Comparison of average AoI, $p=0.4$.}
	\label{res_2}
	\vspace{-0.5cm}
\end{figure}

\begin{figure}[t]
	\centerline{\includegraphics[width=0.9\columnwidth]{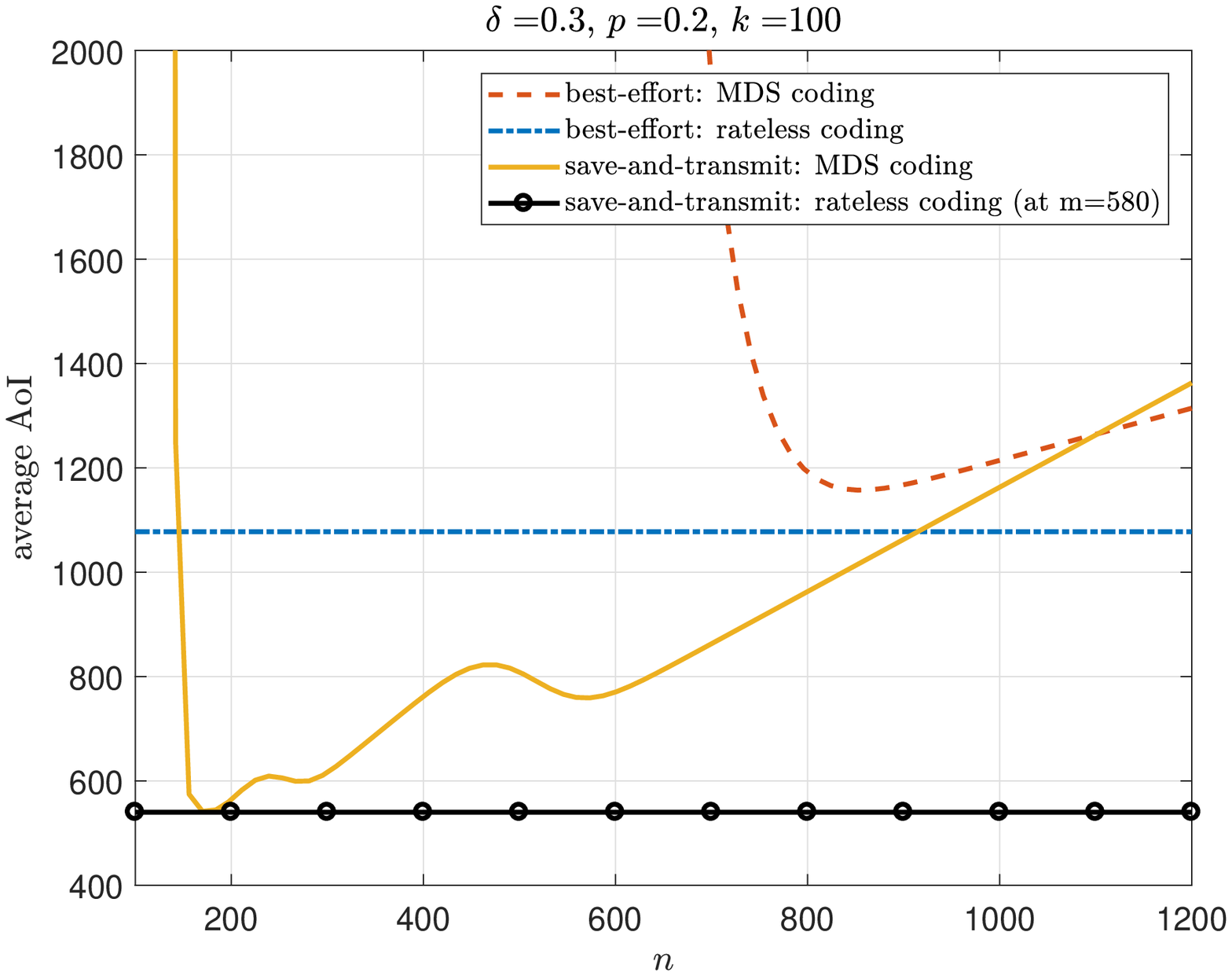}}
	\vspace{-0.5cm}
	\caption{Comparison of average AoI, $p=0.2$.}
	\label{res_3}
	\vspace{-0.6cm}
\end{figure}

In this section, we compare the performances of the proposed schemes. When there is no energy harvesting, i.e., energy arrives with probability $p=1$ at every slot, rateless coding has the best AoI (this mimics the result obtained in \cite{yates2017timely}) and save-and-transmit with MDS coding has the worst performance. The reason that the save-and-transmit with MDS coding has the worst performance is that it requires a saving phase of at least $n$ slots, which is not necessary as the energy arrives at all slots. When the probability of energy arrivals decreases to $p=0.7$, save-and-transmit with MDS coding performs the same as the best-effort rateless coding case, as shown in Fig. \ref{res_1}. Rateless coding with save-and-transmit performs slightly better than all the other policies. As the probability of energy arrival decreases further, save-and-transmit with MDS coding outperforms all the best-effort policies as shown in Fig. \ref{res_2} and Fig. \ref{res_3}. As shown in Fig. \ref{res_3}, the gain becomes significant for low values of $p$. The reason for this is that save-and-transmit eliminates the errors due to energy outage by saving sufficient energy before attempting to transmit. For example, in Fig. \ref{res_3}, for the best-effort scheme, the probability of success in transmitting a symbol is equal to $q=0.2 \times 0.7=0.14$, while if we eliminate the energy outage due to energy harvesting as in save-and-transmit scheme, the success probability for reach symbol will be $0.7$, which is much higher than the best-effort scheme. Rateless coding with save-and-transmit is better than MDS coding with save-and-transmit, because rateless coding with save-and-transmit gives more flexibility for the transmitter to choose just the right saving duration, while in MDS coding case, the transmitter is forced to save for a multiple of $n$ slots.

\bibliographystyle{unsrt}
\bibliography{IEEEabrv,myLibrary}

\end{document}